\documentclass[useAMS,usenatbib,referee]{biom}
\usepackage[pdftex]{graphicx} 

\usepackage{amsmath}

\title[Sieve analysis of non-leaky interventions]{Evaluating the dependence of a non-leaky intervention's partial efficacy on a categorical mark}

\author{Paul T. Edlefsen\emailx{pedlefse@fhcrc.org} \\
Fred Hutchinson Cancer Research Center, Seattle, WA 98109, U.S.A.}

\newcommand*{\argmin}{\operatornamewithlimits{argmin}\limits}

\usepackage{Sweave}
\begin{document}





\pagerange{\pageref{firstpage}--\pageref{lastpage}} 
\volume{00}
\pubyear{2013}
\artmonth{December}


\doi{10.1111/j.1541-0420.2005.00454.x}


\label{firstpage}

\begin{abstract}
  We address discrete-marks survival analysis, also known as
  categorical sieve analysis, for a setting of a randomized
  placebo-controlled treatment intervention to prevent infection by a
  pathogen to which multiple exposures are possible, with a finite
  number of types of ``failure''.  In particular, we address the case
  of interventions that are partially efficacious due to a combination
  of failure-type-dependent efficacy and subject-dependent efficacy,
  for an intervention that is ``non-leaky'' (where ``leaky''
  interventions are those for which each exposure event has a chance
  of resulting in a ``failure'' outcome, so multiple exposures to
  pathogens of a single type increase the chance of failure).  We
  introduce the notion of some-or-none interventions, which are
  completely effective only against some of the failure types, and are
  completely ineffective against the others.  Under conditions of no
  intervention-induced failures, we introduce a framework and Bayesian
  and frequentist methods to detect and quantify the extent to which
  an intervention's partial efficacy is attributable to uneven
  efficacy across the failure types rather than to incomplete ``take''
  of the intervention. These new methods provide more power than existing methods to detect
  sieve effects when the conditions hold.  We
  demonstrate the new framework and methods with simulation results
  and new analyses of genomic signatures of HIV-1 vaccine effects in
  the STEP and RV144 vaccine efficacy trials.

\end{abstract}

%



\maketitle


%

\DefineVerbatimEnvironment{Sinput}{Verbatim} {xleftmargin=2em}
\DefineVerbatimEnvironment{Soutput}{Verbatim}{xleftmargin=2em}
\DefineVerbatimEnvironment{Scode}{Verbatim}{xleftmargin=2em}
\fvset{listparameters={\setlength{\topsep}{0pt}}}
\renewenvironment{Schunk}{\vspace{\topsep}}{\vspace{\topsep}}

\section{Introduction}
\label{s:intro}

%

Despite advances in both treatment and prevention, the global HIV/AIDS
epidemic continues to levy a heavy toll on millions of HIV-infected
individuals and those who love them. With a global prevalence of over
33 million, and an estimated 14\% prevalence
among women and men aged 15-49 in some subsaharan African states
where non-vaccine interventions are difficult to scale to a population
level, there is great need for an effective
public health vaccine \citep{HIVPrevalence}.

There is no licensed vaccine to prevent infection by HIV.  While some vaccines have been
relatively easy to develop, creating effective vaccines against HIV has
proven more challenging.  In response to the slow pace of
the traditional strategy for vaccine development, organizations such
as the Global HIV-1 Vaccine Enterprise are advocating for more iterative
strategies, in which even ``failed'' clinical trials nevertheless provide maximal
insight into mechanisms of whatever partial protection may have been
conferred to some subset of vaccine recipients \citep{GlobalVaccineEnterprise2010}.

One critical question that arises in such an approach is,
in the aftermath of a randomized clinical trial to evaluate a new
treatment intervention, to what extent should observed differences in
apparent efficacy across outcome types be attributed to actual imbalanced
efficacy across those types, as opposed to imbalanced uptake across
subjects, or to just poor efficacy overall?  This is non-trivial because other
factors may also lead to a difference in observed failure rates across
types, other than differential efficacy of the intervention to prevent
type-specific failures.

A sieve analysis investigates an interventions's efficacy as a function of specific features of
the outcome \citep{edlefsen2013current}.  A recent
example comes from a sieve analysis of the RV144 HIV-1 vaccine
efficacy trial.  The RV144 vaccine regimen was partially
efficacious at reducing HIV-1 acquisition risk among healthy low-risk
Thai volunteers, with an estimated intervention efficacy (IE) of 31\%
versus placebo \citep{Ngarm}.  A subsequent sieve analysis identified
that this efficacy significantly differed against viruses of two
types that are differentiated by a viral genome feature at an immunologically
relevant locus \citep{v2sieve}.  This result concords with other
evidence that the vaccine induced antibodies that target a specific
region (called V2) of the viral envelope \citep{haynes2012immune}, and together
these results are influencing ongoing and planned HIV-1 vaccine
trials.

Sieve analysis can also provide valuable insight into an
intervention's effect in the context of an intervention with no
overall efficacy.  An example comes from the STEP HIV-1 vaccine
efficacy trial,
which was halted early for efficacy futility \citep{Buchbinder}.
STEP evaluated a vaccine candidate in high-risk
study participants in the Americas.  At its early conclusion, the
estimated hazard ratio of infection (vaccine vs placebo) was about 1.5.
A sieve identified vaccine-driven selective pressure at an
immunologically-relevant HIV-1 locus, indicating that although it did not
reduce the overall acquisition rate, the vaccine did induce an immune
response that affected the distribution of observed viruses, possibly
by altering the dynamics of post-infection viral evolution
\citep{Rolland}.

Sieve analysis is also known as ``mark-specific efficacy analysis'',
where the general term ``mark'' refers to features of the failure (as
opposed to covariates of the subject).  Sieve analysis methods have
been developed for a variety of types of marks \citep[e.g.][]{GSA,
  GMS, juraska2013mark}.  Common example features
include pathogen serotypes, and quantitative measures of pathogen susceptibility to
neutralization.  The framework that we introduce in this article
addresses finite-category nominal marks.

Peter Gilbert
enumerated several conditions that are required to yield unbiased
estimates of mark-specific intervention efficacy for the case of
a ``leaky'' intervention \citep{GSA, gilbert2001interpretability}.  The notion of a ``leaky'' intervention was
introduced by \cite{struchiner1990behaviour} in contrast with an ``all-or-none''
intervention \citep{halloran1991direct} for a setting in which failure types are not
differentiated, and has been extensively developed for applications in
that setting \citep{halloran2010design}.  In the former case, each of a subject's
potentially-multiple exposures has some chance of being thwarted by
the effect of the intervention; in the latter case, each subject is
either a responder (and therefore protected if exposed) or not.
In vaccine trials, responders are said to have ``take'' of the vaccine.

Peter Gilbert argued and showed through simulation studies that his
conditions for unbiased sieve analysis are violated whenever some
subjects have no ``take'' \citep{gilbert2000comparison}.  We introduce
a new variant on the ``leaky'' vs ``all-or-none'' dichotomy: the
``sievey-not-leaky'' intervention, which exhibits sieve effects but
not per-event stochastic leaks.  We call such an intervention
``some-or-none'', as it completely protects subjects with ``take'' -
but only against failures with marks that fall in an
``intervention-targeted'' subset.  We provide analogous conditions for
unbiased assessment of sieve effects in this non-leaky setting, and
introduce the ``sievey-not-leaky'' modeling framework for evaluating
a some-or-none intervention under the ``no-harm'' condition (that the
intervention can prevent would-be-failures but cannot induce new
failures or increase susceptibility).

We also provide a new perspective on the 
``Model-Based Sieve'' (MBS) method that was employed in the RV144
sieve analysis \citep{v2sieve}, and show that it can be understood
either as an approximation to the some-or-none intervention model or
as a model of the effects of the intervention on post-infection variation.

The scope of the present work is limited to categorical
representations of all outcomes.  Future work
incorporating time-to-event data into the framework is likely to
improve efficiency of the methods.  Note too that this representation
necessitates excluding subjects lost-to-followup before the end of the
trial (assuming conditions of uninformative missingness), who may
contribute information relevant to estimation of intervention efficacy
but would not (in this framework) contribute to the estimation of
relative failure rates across failure types.

In the Results section we present new analyses of the genomic
sequences of the HIV-1 viruses that infected participants of the RV144
and STEP trials, in which the mark types are the amino acids (AAs)
found in the genome of each virus at a particular locus of interest.
In that setting, one particular AA is unique because it was included
in the vaccine (so we expect that sieve effects will reflect greatest
efficacy against this targeted category) but the other AAs are
unordered.  We address the general problem of sieve analysis for
unordered categories, and are particularly motivated by settings in
which some categories are targets of the intervention and others are
not.

The remainder of this article is organized as follows.  First we
describe the ``sievey-not-leaky'' (SNL) modeling framework.  Although
we assume that the intervention does not induce new failures, sieve
effects may lead to type-specific increases in the observed failure
rates due to replacement of avoided failures by failures of other
types.  We discuss the potentially confounding effects of such
``replacement causes'' on efforts to detect sieve effects, and
describe conditions under which observed changes to failure type
distributions are attributable to sievey-not-leaky sieve effects.  We
outline methods to test these conditions and present Bayesian and
frequentist methods to detect sieve effects when the conditions hold.
To our knowledge, this is the first application of Bayesian
methodology to the problem of sieve analysis.

\section{Sievey-not-leaky modeling of intervention efficacy}

\subsection{Notation}

As we introduce the SNL framework and some-or-none models, we will use
the following notational conventions.  All vector-valued parameters
are in bold and parameter vectors 
are named consistently so that their dimensionality and role is
apparent.  Lowercase subscripts convey either the main role of the parameter, the treatment
group to which it corresponds, or an index into the (vector-valued)
parameter.  Table~\ref{t:notation} describes the relevant symbols.

\begin{table}[ht]
  \caption{Notation symbols and their meanings}
\begin{center}
\begin{tabular}{r l l l l l }
  \hline
& & & & role & index\\ 
category & symbol & meaning & range & subscripts & subscripts\\ 
  \hline
& $J$ & number failure categories & $\mathcal{N}^+$ & & \\         
& $g$ & number targeted categories & $1 \dotsc J$ & & \\          
& $R$ & set of response categories & $R = \{0,1\}$ & & \\           
& $\tau$ & (time) duration of the trial & $\mathcal{R}^+$ & & \\            
initial & $z$ & treatment group & $\{ \hbox{p}, \hbox{v} \}$ & & $i$ \\         
initial & $n$ & outcome count & $\mathcal{N}$ & $p$, $v$, $c$ & $j$ \\         
initial & $p$ & probability & $[0,1]$ & $r$, $s$ & \\  
initial & $r$ & ``take'' response indicator & $R$ & & $i$ \\ 
initial & $\bm{Y}$ & outcomes for each subject & $0 \dotsc J$ & $p$, $v$, $c$ & $i$ \\       
initial & $\bm{n}$ & count simplex & $[0,1]^{J+1}$ & $p$, $v$, $c$ & $j$ \\       
initial & $\bm{r}$ & probability simplex & $[0,1]^{J+1}$ & $p$, $v$, $c$ & $j\ge0$ \\     
initial & $\bm{p}$ & probability simplex & $[0,1]^{J}$ & $p$, $v$, $c$ & $j>0$ \\    
initial & $\bm{q}$ & probability simplex & $[0,1]^{J-g}$ & $p$, $v$, $c$ & $j>g$ \\   
initial & $\bm{t_i}$ & exposure times for subject $i$ & $(0, \tau)$ & & $j >0$ \\            
initial & $\bm{\omega_r}$ & failure avoidance rates for response $r$ & $[0,1]$ & & $j > 0$ \\
subscript & $i$ & index for subjects & $1 \dotsc n$ & & \\ 
subscript & $j$ & index for outcome types & $0 \dotsc J$ & & \\ 
subscript & $p$ & control treatment group & & & \\ 
subscript & $v$ & intervention treatment group & & & \\ 
subscript & $c$ & counterfactual for intervention group & & & \\ 
special & $t$ & $p_t \equiv Pr( \hbox{``take'' response} )$ & & & \\  
special & $2$ & $p_2 \equiv Pr( \hbox{avoided failure is replaced} )$& & & \\   
special & $s$ & $p_s \equiv$ the ``sieve effect strength''& & & \\ 
  \hline
\end{tabular}
\end{center}
 \label{t:notation}
\end{table}

We begin with (categorical, completely observed) outcome data $\bm{Y}$
for all $n = n_v + n_p$ subjects of a (placebo-)controlled randomized
trial, with $\bm{Y_v}$ denoting the outcomes for just the $n_v$
intervention recipients (that is, for those subjects for whom $z_i =
v$).  Outcome types $j$ include the special value $0$ to indicate
non-failure, and the other (failure) types are labeled such that $1$
through $g$ are
failure types that are targeted by the $g$-or-none intervention. For
instance $n_{p0}$ of the $n_p \equiv \sum_{j=0}^{J} n_{pj}$ control
recipients had not failed by the end of the trial, and $n_{v1}$ of the
$n_v \equiv \sum_{j=0}^{J} n_{vj}$ treatment intervention recipients
failed with failure type $1$.
 
We let ``c'' (as a subscript) represent the counterfactual
values of what would be seen in the treatment group if they had not
received the treatment.  So for instance we define the counterfactual
total number of subjects $n_c$ as
$\sum_{j=0}^{J} n_{cj}$, which is equivalent to $n_v$ since by
assumption the intervention does not affect the total number of
subjects (just which subjects fail, and their corresponding failure
types).

Uppercase subscripts (``G'') in the index position indicate a scalar
sum over indices $1 \dotsc g$.  For example, $r_{cG} = \sum_{j=0}^g{p_{cj}}$
is the probability that a treatment-intervention recipient would have
a targeted-type failure if that subject had not received the
intervention.  Parameters with initial ``r'' refer to outcome
proportions, with ``p'' refer to failure proportions, and with ``q''
refer to non-targeted-type proportions; so for example $r_{pJ} \equiv
\frac{n_{pJ}}{n_p}$, vs $p_{pJ} \equiv \frac{n_{pJ}}{n_p - n_{p0}}$,
vs $q_{pJ} \equiv \frac{n_{pJ}}{n_p - ( n_{p0} + n_{pG})}$.  Note that
we also use $\bm{q}$ without a treatment group subscript to indicate
the distribution among non-targeted categories of replacement
failures, as discussed below.

To be consistent with the literature on discrete marks sieve analysis,
we parameterize our approach to comparing some-or-none to all-or-none
models in terms of the failure type frequencies $p_{vj}$ and $p_{pj}$.  Although
when optimizing parameters (for maximum-likelihood estimation) we
may employ the logistic transform for more convenient search, in this
article we represent the category probabilities directly in their own
simplex scale (rather than the logistic scale).  As all of the model
parameters index either a binomial or multinomial distribution family,
when being Bayesian we use conjugate beta or Dirichlet priors on this
simplex scale.

\subsection{No-harm, would-be-first failure processes, and replacement
failures}

In accord with the definition used in \cite{GSA}, we define
mark-specific intervention efficacy in terms of ``per-exposure''
probabilities of infection (failure), given one exposure event.  Since
we are not able to observe exposures, and since a single subject may
experience zero or multiple exposure events, we are not generally able to
directly estimate these per-exposure probabilities \citep{rhodes1996counting}.  \cite{GSA} showed
that under certain (leaky) conditions, relative risks and odds ratios of these
probabilities are directly estimable.

Departing somewhat from these and other articles' definitions of 
``exposure'', in this article (and in the
framework here introduced) we simplify the setting to assume that
``exposure'' is defined as ``an exposure that would, in the absence of
intervention, result in a failure''.  Thus the per-exposure probability
of failure is $1$ for placebo recipients (and for intervention
recipients in the counterfactual case) for any type of failure, and
variations of failure rates across types for these subjects are due
solely to variations in exposure distributions across the failure types.  Note
that this definition precludes the possibility that an intervention-receiving
subject would experience a new failure that would not be experienced
in the counterfactual case.  This is the ``no-harm'' assumption.

However, since an intervention-receiving subject might avoid (due to
the intervention) the failure that would have affected him in the
counterfactual case, and is thus potentially susceptible to subsequent
exposures, there may be ``replacement failures'' affecting intervention recipients
that would never have been observed in the counterfactual case (since
a subject's observed failure is defined as her \textit{first} exposure that is
not avoided).

When there are $J$ failure types, this exposure model is equivalent to a $J$
competing risks model in which only the time of the ``winning'' exposure
process is observed, where ``winning'' additionally requires that the exposure is
not avoided due to the intervention.  Concretely, we represent the $J$ competing risks
for each subject $i$ with $J$ simultaneous (and possibly
jointly-distributed) continuous-time Markov counting processes
$N_{ij}(t)$, $j \in 1, \dotsc, J$, where at time $t$ the state of
process $N_{ij}(t)$ is a natural number indicating the number of
potential failure events of type $j$ that subject $i$ would experience
by time $t$ if all such potential failures up to time $t$ were avoided
(that is, they did not become actual failures) due to the
intervention.

This is subtle, but it is important to note that
$N_{ij}(t)$ is not the same as the counting process of subject $i$'s
actual failures, $N'_{ij}(t)$, in which the probability of a
subsequent actual failure, given all previous failures, reflects a
potential change in the subject's susceptibility due to the previous
failures.  Instead, process $N_{ij}(t)$ always reflects the surviving
subject's chance of experiencing an exposure, and the state of this
process represents the number of previous would-be-failures that were
avoided due to the intervention effect.


This formulation assumes independence between the failure processes of
every pair of subjects.  Note, though, that we assume nothing
particular about the failure processes $N_{ij}(\cdot)$ for any one
subject $i$: they may be arbitrarily jointly distributed across the
$J$ types.  In particular we do not assume that they are memoryless or
of any exponential family.

By construction, a subject who does not receive the intervention will
never experience an intervention-induced failure avoidance.  We
nevertheless define these ``would-be-first failure processes'' for all
subjects; any control-recipient subject for whom any of
the $J$ processes $N_{ij}(\tau)$ is non-zero at the end of the trial
will have $Y_i > 0$ (since any subject $i$ with $z_i = p$ will
experience a failure of type $j$ by time $\tau$ whenever $N_{ij}(\tau)
> 0$).  Since the failure type is defined as the mark of the first
failure that actually occurs during the trial, we can define
control-recipient failure types in terms of the would-be-first failure
processes as $Y_i = \argmin_j t_{ij}$, where $t_{ij} \equiv \argmin_t
{N_{ij}(t) I\left(N_{ij}(t) > 0\right)}$, or $Y_i = 0$ if all $J$
times $t_{ij}$ are greater than $\tau$.

\subsection{Response types $R$}

There may be a subgroup of treatment intervention recipients for whom
the treatment did not have any relevant effect.  In general we allow
for a subject-specific latent (unmeasured) response variable $r_i \in
R$ to represent subject $i$'s response to the intervention (eg. immune
response to vaccination), with the constraint that a unique value in
the set $R$ represents ``no response'' ($r_i = 0$), and that all
control recipients (subjects for whom $z_i = p$) have $r_i = 0$.  We
let $p_t$ represent the rate of non-zero responses among intervention
recipients (understood as an expected value over possibly varying
subject-specific response rates $p_{ti}$, so $p_t \equiv E_i[ Pr( r_i
\ne 0 | z_i = v ) ]$).

For simplicity of presentation we here consider only binary responses,
$R \equiv \{ 0, 1 \}$, where subjects for whom $r_i = 1$ are the
subjects with ``take'', whose failures are potentially influenced by the
intervention.  We henceforth use the terms ``responders'' and those
with ``take'' interchangeably.  In a setting such as genome-scan sieve
analysis, in which multiple features (genomic loci) are evaluated, the
definition of ``take'' might be feature-specific, as in, ``the subject
experienced vaccine-induced immune responses that target this locus.''
We do not represent ``take'' on a per-failure-type basis. In the Discussion section (below) we address the
extension of this ``dichotomous-take'' framework to a setting with
multiple response types, which includes the special case of models
with independent per-failure-type response rates.

\subsection{Failure avoidance rate vectors $\{ \bm{\omega_r} : r \in R\}$, some-or-none interventions}
 
For any particular response type $r \in R$ let there be a vector
$\bm{\omega_r}$ of $J$ values, with the value at index $j$ (denoted
$\omega_{rj}$) representing the per-event (aka ``per-exposure'')
probability that a subject with treatment response $r$ avoids an event
that would, in the absence of the intervention, have caused a failure
of type $j$.  We assume that these avoidance rates are constant in
time, leaving
to future work the interesting extension to time-varying efficacy.

The setting of multiple failure types leads to an interesting
variation on the ``all-or-none'' versus ``leaky'' treatment
intervention dichotomy.  By definition, treatment interventions
exhibiting sieve effects can never be ``all-or-none'' (which would
have equal and perfect avoidance for all takers regardless of failure
type); but they could be ``some-or-none'' (or ``one-or-none'') if for
all $r$ and $j$, $\omega_{rj} \in \{ 0, 1 \}$ but there exists some
pair of types $(j, j')$ for which $\omega_{rj} \ne
\omega_{rj'}$.  Such an intervention induces (in takers) perfect
avoidance of some but not all failure types (at any particular time).
Consider, for concreteness, a dichotomous-take one-or-none
intervention with $\bm{\omega_1} = ( 1, 0, \dots, 0 )$.  We do not consider this a ``leaky'' intervention,
since all of the IE is determined through the take indicator $r_i$,
with no per-event stochastic filtering.



When it is not known a priori which type(s) might be the targets of
the intervention, the one-or-none and some-or-none models can be used
in concert to compare different hypotheses regarding the nature of a
particular sieve effect.  When the SNL sieve conditions hold (see below),
and particularly under the condition that the intervention is non-leaky, these
models reflect all possible sieve effects for which avoidance rates
that are non-zero are the same for all types at any given time.  In a
Bayesian setting, for instance, we can compute the posterior
probability of every possible some-or-none SNL model and
compare these to the posterior probability of the all-or-none null
model.

We can also compare particular pairs of models using their posterior
probabilities, or by using likelihood ratio
tests or Bayes factors.  For instance we can compare the all-or-none
sieve null model (in which treatment recipient relative failure type
rates are the same as the counterfactual distribution, which we
estimate from the control-recipient data) to the particular
one-or-none model in which takers avoid all type-$1$ failures. In the
absence of replacement failures, the only effect of the one-or-none
intervention is to reduce the type-$1$ failure rate for ``takers'' by
$p_t$, so this test compares a model reducing just type-$1$ rates by
$p_t$ to a model reducing all rates by $p_t$.

\subsection{The cause replacement rate $p_2$}


We can make the notion of replacement failures precise in the specific
context of some-or-none interventions with target types $1, \dotsc, g$
($\omega_{1j} = I( 1 \le j \le g )$).  In this context, replacement
causes imply that $( n_{v0} + n_{vG} ) < ( n_{c0} + n_{cG} )$.  Put
another way, replacement failures cause some treated subjects to fail
with a non-targeted failure type who would (in the absence of
treatment) have failed with a targeted failure type, so
$\sum_{j=(g+1)}^{J} n_{vj} > \sum_{j=(g+1)}^{J} n_{cj}$.

If we assume that a
dichotomous-take some-or-none intervention is completely effective (for subjects with
relevant responses) at preventing failures of types $1$ through $g$,
then the expected number of treatment recipients who fail with each of
these types is $E[n_{vj}] = n_{cj} (1 - p_t)$.
Let $p_2$ be the probability that a
subject who would have failed by a targeted failure type has a
replacement failure during the trial.  Then the expected total number
of uninfected treatment recipients is $E[n_{v0}] = n_{c0} + n_{cG} p_t (
1 - p_2 )$.  The remaining subjects with
intervention-induced avoidance of their targeted failures go on to
fail anyway by one
of the remaining types.  Let $q_j$ be the conditional probability that the
replacement failure is of type $j \in (g+1), \dotsc, J$, with $\sum_j q_j
\equiv 1$, so that the product $p_2 q_j$ is
the probability that the replacement failure occurs before the end of
the trial and is of type $j$. Then the expected total number of failures of type $j > g$
among treatment recipients will be $E[n_{vj}] = n_{cj} + n_{vG} p_t p_2
q_j$.

The cause replacement rate $p_2$ is the proportion of the $n_{cG}$ subjects (who in the
absence of the treatment intervention would have had targeted-type
failures during the trial before any non-targeted-type failures) with
a relevant treatment response ($r_i = 1$) for whom at least one of the
non-targeted failure processes also fails during the trial.  In our
notation, $p_2 = Pr( \exists j > g : t_{ij'} < t_{ij} \le \tau | r_i =
1, \min_{j \le g}{(t_{ij'})} \le \tau )$, where $t_{ij} \equiv \min_t
{(N_{ij}(t) = 1)}$.

Note that the conditional distribution among the other categories (those not
targeted) will be unaffected by a one-or-none or some-or-none
intervention unless there are replacement failures.  That is, if $p_2
= 0$, then $q_{vj} = q_{cj}$ for all $j > g$.  If instead there are replacement
failures, the conditional probabilities of the off-target failure
types will increase to some degree, since $p_{vj} = \frac{p_{cj} + p_{cG} p_t p_2 q_j}{1 - I_E}$.


\subsection{SNL sieve conditions}
As we have discussed, we assume conditions of no harm and no
stochastic leaks. Together, these conditions ensure that replacement failures
occur only in the presence of sieve effects, never under the
all-or-none null.  In this context we still cannot guarantee that
naive estimates of mark-specific efficacy using relative rates of
failure will be unbiased, since replacement failures in the presence
of sieve effects may arbitrarily and unpredictably alter the
$\bm{p_v}$ for the untargeted types.  However since there are a
limited number of some-or-none models, if the conditions hold then the
problem of identifying a sieve effect becomes the problem of
identifying which of those models holds best.

Two additional conditions are required. First, we assume time-constancy (over
the duration of the trial) in
all of the parameters, notably the overall $I_E$, the two
intervention response parameters $p_t$ and $\bf{\omega_1}$, the
counterfactual failure type distribution $\bm{p_c}$, and the
cause replacement rate $p_2$ and distribution $\bm{q}$.  As \cite{gilbert2001interpretability}
discussed for the leaky-intervention sieve analysis framework, it may
be possible in future efforts to relax some of these time-constancy constraints
either by reinterpreting the model parameters (as averages over time) or by augmenting the
model.

Second, as well as requiring between-subjects independence of all
outcomes, we require independence between a subject's response $r_i$
to the intervention and the relative failure rates of his would-be-first failure process
distributions.  This is (all that is) required to avoid the bias
engendered if subjects with ``take'' tend to also be subjects with a particular
type of failure.

\subsection{The sieve effect strength $p_s$}
\label{s:p_s}

In our mathematical models of some-or-none interventions, we do not explicitly represent the take rate $p_t$ or the cause replacement rate $p_2$.
We now introduce the primary parameter of these models, $p_s \equiv \frac{p_t ( 1 - p_{cG}( 1 -
  p_2 ) )}{1 - p_t p_{cG} (1 - p_2 )}$, which subsumes the role of
those other two parameters.  As detailed in Web Appendix A, the ``sieve effect strength'' $p_s$ expresses the magnitude of a sieve
effect as the proportion of those subjects who would not be protected by
the some-or-none intervention who would instead be protected if the
intervention were an all-or-none intervention.  That is,
it reflects the extent to which an intervention's partial efficacy can
be attributed to its sieve effect as opposed to its incomplete rate of
take: the partial efficacy of a some-or-none intervention with $p_s =
0$ has nothing to do with sieving, whereas the partial efficacy of a some-or-none
intervention with $p_s = 1$ is completely attributable to its sieve effect.

Note that the SNL framework implies the constraint $I_E \le p_t p_{cG}$.
As elucidated in Web Appendix B, this constraint can be expressed
solely in terms of the some-or-none model parameters $I_E$, $p_{cG}$,
amd $p_s$.  See Web Appendix C for a mathematical representation of the some-or-none model
likelihood in the SNL framework.

\subsection{Replacement-only models and $p_s$}

If $I_E = 0$ (``replacement-only''), the sieve effect strength parameter $p_s$ becomes
equivalent to the take rate $p_t$, and all avoided failures of the
targeted types are replaced (that is, $p_2 = 1$). Then the conditional
probabilities of the off-target failure types ($j > g$) will be
$p_{vj} = p_{cj} + p_{cG} p_s q_j$, and for the targeted types $j \in
1, \dotsc, g$, $p_{vj} = p_{cj}( 1 - p_s )$.

The replacement-only all-or-none model and the
replacement-only some-or-none model constrained to $p_s = 0$ have the same conditional likelihood of the
failure categories $p_{vj}$ as the counterfactual $p_{cj}$.  Thus the
likelihood is
mathematically equivalent between them, which makes the likelihood ratio test
comparing the all-or-none to the some-or-none model a candidate for
Wilks's theorem (of an asymptotic chi-squared distribution with one
degree of freedom, reflecting model equivalence when the parameter
$p_s$ is zero), but only when $I_E = 0$. In the Results section, we
demonstrate in simulations that the chi-squared(1) null distribution holds well for
replacement-only scenarios.  We also show that a chi-squared(1) null controls size
in our numerical experiments even for cases with $I_E > 0$.

\subsection{MBS, replacement-only, and insert-only models}
\label{s:MBS}

We previously introduced an approach, ``Model-Based Sieve'' (MBS)
\citep{v2sieve}, that is closely related to replacement-only
one-or-none models in the SNL framework.  Briefly, the MBS method compares the probability
of the intervention-recipient failures given a ``null'' multinomial model
with parameters estimated as being proportional to the observed
placebo-recipient frequencies (plus pseudocounts of $1/J$ per category)
to the ``alternative'' model
probability that is computed as the expected value (over an
``insert-only'' indicator parameter, $i \sim \hbox{Bernoulli}(0.5)$,
and a ``sieve effect strength'' parameter, $p_s \sim
\hbox{Uniform}(0,1)$) of the probability of the intervention-recipient
sequences given a multinomial model in which the probability of the
targeted type is multiplied by $(1-p_s)$ and the removed mass is reallocated either proportionally (if $i$ is 1) or uniformly (if $i$ is 0) among the remaining categories.

The MBS approach employs a
special case of the replacement-only one-or-none model. As
such it is relevant for the evaluation of post-acquisition effects, in
which failure types may be redistributed due to the intervention, but
any subject who would have failed in the absence of the intervention
also fails in the presence of it.  The MBS method compares (using Bayes factor) the all-or-none null
model to an even mixture of two one-or-none models (both of which
assume no intervention efficacy).  Unlike the other some-or-none
SNL models, MBS does not average over a prior distribution for
the replacement distribution $\bm{q}$ (nor does it compute an MLE). The two components of the mixture
model differ in their choice of $\bm{q}$:
one model uses an evenly-distributed $\bm{q}$ replacement vector
and the other (called the ``insert-only'' model) uses $q_j \equiv
\frac{p_{cj}}{1 - p_{cG}} = q_{cj}$.

The mixture model employed by the MBS approach effectively combines two
models that vary in whether the effect is restricted to the targeted
types or also affects off-target types. In general when $I_E = 0$ (``replacement-only''),
$q_{vj} = \frac{p_{cj} + p_{cG} p_s q_j}{1 - p_{vG}} = \frac{p_{cj} +
  p_{cG} p_s q_j}{1 - p_{cG}(1-p_s)} = ( q_{cj} + \frac{p_{cG} p_s}{1
  - p_{cG}} q_j ) \frac{1 - p_{cG}}{1 - p_{cG}(1-p_s)}$.  Thus under
the additional condition of the insert-only model (that $q_j =
q_{cj}$), $q_{vj} = ( q_{cj} ( 1 + \frac{p_{cG} p_s}{1 - p_{cG}} ) )
\frac{1 - p_{cG}}{1 - p_{cG}(1-p_s)}$, and since $( 1 + \frac{p_{cG}
  p_s}{1 - p_{cG}} ) = \frac{ 1 - p_{cG}}{1 - p_{cG}(1 - p_s)}$,
$q_{vj} = q_{cj}$.  Therefore the effect of a some-or-none treatment under
the insert-only model is entirely restricted to reducing relative
failure rates against the targeted types (there is no effect on the
conditional distributions of the remaining types), a fact that
motivated the name of this model (``insert-only'').  The name reflects
the common situation in efficacy trials where a vaccine only protects
against pathogen serotypes included in the vaccine
immunogen (``insert'').


\subsection{Two-phase modeling}
Bayes factors for the MBS method are calculated using only the
intervention-recipient data, after first using the placebo-recipient
data to update priors on the counterfactual failure type distribution
$\bm{p_c}$.  We call this approach ``two-phase'' because the
placebo-recipient data is used in a first phase of analysis (to update
priors) and then only vaccine-recipient data is used in a second phase.


Although in a Bayesian setting it is natural to model the parameters
hierarchically to incorporate prior uncertainty about them,
non-hierarchical approaches are also reasonable when parameter
uncertainty is small or as an approximation to a fully hierarchical
approach.  The MBS method is a ``hierarchical two-phase'' method, in
that the Bayes factors are computed for the conditional distribution
of the intervention-recipient failure count data alone, after first
using the placebo-recipient data to update the prior distribution of
$\bm{p_c}$.

%

\section{Results}
We created reference implementations (in R) for hypothesis tests
based on log-likelihood ratios and Bayes factors (for comparing
all-or-none to individual some-or-none models).  Using these
implementations, we use a simulation study to show that p-values
approximated using chi-squared distributions (with one degree of
freedom) control size in the context of a small-sample example
setting, even for non-replacement-only scenarios.  We show that the simple decision rule to reject the null
hypothesis (of no sieve effect) whenever the Bayes factor favors the
alternative hypothesis (of a particular some-or-none sieve effect)
yields a powerful hypothesis testing procedure in the frequentist
sense (and in our simulations controls size fairly well), though as expected the benefits of the Bayesian
approach are most relevant to inference (such as investigating the
posterior distribution of $p_s$) than to frequentist hypothesis
testing; we also provide a comparison of Bayesian approaches
using the area under receiver operating characteristic
(ROC) curves (which consider all possible prior model odds).  We show that a Bayesian-frequentist hybrid approach of
determining a null distribution for the Bayes factor through
permutation of class labels retains much of the power of the direct
Bayes factor approach (while guaranteeing size control under the
permuted null).  We also demonstrate the power
of these methods in comparison existing methods.  Finally, we perform a new
sieve analysis of two sites found previously to exhibit sieve effects
in actual HIV-1 vaccine trial infection data.  Our analyses support the sieve
effects at these sights and provide new analyses of the ``sieve effect
strengths'' of these effects.

\subsection{The simulation study}
\label{s:simulation}
For the simulation study we considered $J = 5$ failure types and $g =
1$ targeted type.  We simulated outcome counts for models with a
$10\%$ failure rate among $n_p = 1000$ placebo recipients ($r_{c0} =
.9$).  Mimicking the typically-geometric distribution across amino
acids observed at a typical locus in our HIV-1 genome-scanning
applications, we set the simulated data $\bm{p_c}$ vector given by
$\bm{p_c} \equiv (0.815, 0.171, 0.0135, 0.0005,
0.000006)$.  When simulating from an insert-only model, we set the
replacement distribution $\bm{q}$ proportional to the non-targeted
type elements of $\bm{p_c}$, and for non-insert-only models we set
$\bm{q}$ to be uniform ($\bm{q} \propto 1$).  We also simulated $n_v = 1000$
intervention-recipient outcomes from the one-or-none (or all-or-none)
model.

We generated $1000$ simulated data sets for each of eleven scenarios.
The constraints of the model prohibit simulation of the case
with $I_E = 0.5, p_s = 0.15$ 
so we use $I_E = 0.2$ for that scenario instead.
Otherwise, these
cases cover all non-insert-only scenarios with
$I_E \in \{ 0, 0.50 \}$
and $p_s
\in \{0.15, 0.25, 0.50 \}$, as well as the replacement-only insert-only cases
(in which $\bm{q} \propto \bm{p_c}$).  We omit the other
insert-only scenarios, which as discussed in Web Appendix D differ only in their
value of $I_E$ from these replacement-only insert-only scenarios.

To evaluate the size of the tests, we include two null distributions:
the all-or-none null with $I_E=0.5$ exhibits no sieve effects because
the data are simulated without such effects; the permuted one-or-none
null is created by simulating 1000 datasets from the one-or-none
scenario, then independently permuting treatment labels of each one.

Note that while the total sample size in all of these scenarios is
fixed at $n_v = n_p = 1000$, the effective sample size for estimating
all parameters other than $I_E$ depends on the value of $I_E$.  Thus
for the simulations in which $I_E = 0$, about $100$ vaccine
recipients become infected, whereas for $I_E = 0.20$, about $80$ do,
and for $I_E = 0.50$, about $50$ vaccine recipients become infected in
each of the $1000$ randomly generated datasets per scenario.  This
cautions against comparisons of power across scenarios with different values of $I_E$.

These simulated datasets were used to evaluate several variations of
the sievy-not-leaky methods, as well as the ``GWJ'' nonparametric
(permutation-based) t-test \citep[statistic $Z^A_2$ in][]{GWJ}, and a simple
Fisher's exact test of the failure tables (including only counts of
failures).  Of the many possible SNL approaches, here we evaluate two
fully frequentist approaches that compare minus two times the
log-likelihood ratio (LLR) to a chi-squared with one degree of freedom
(``1phase'' and ``2phase''), rejecting the null hypothesis when the
test statistic exceeds the $95^{\hbox{th}}$ percentile of the analytic null
distribution.  We also evaluate three Bayesian procedures
(``BF1ph'', ``BF2ph'', ``MBS-BF'') based on comparing the estimated
Bayes factors (BFs) to $1$ (``rejecting'' whenever BF > 1), and
permuted-null versions of each (``BF1ph-perm'', ``BF2ph-perm'', and
``MBS'', respectively).







\begin{figure*}
\label{f:heatmap}
\begin{center}
\includegraphics{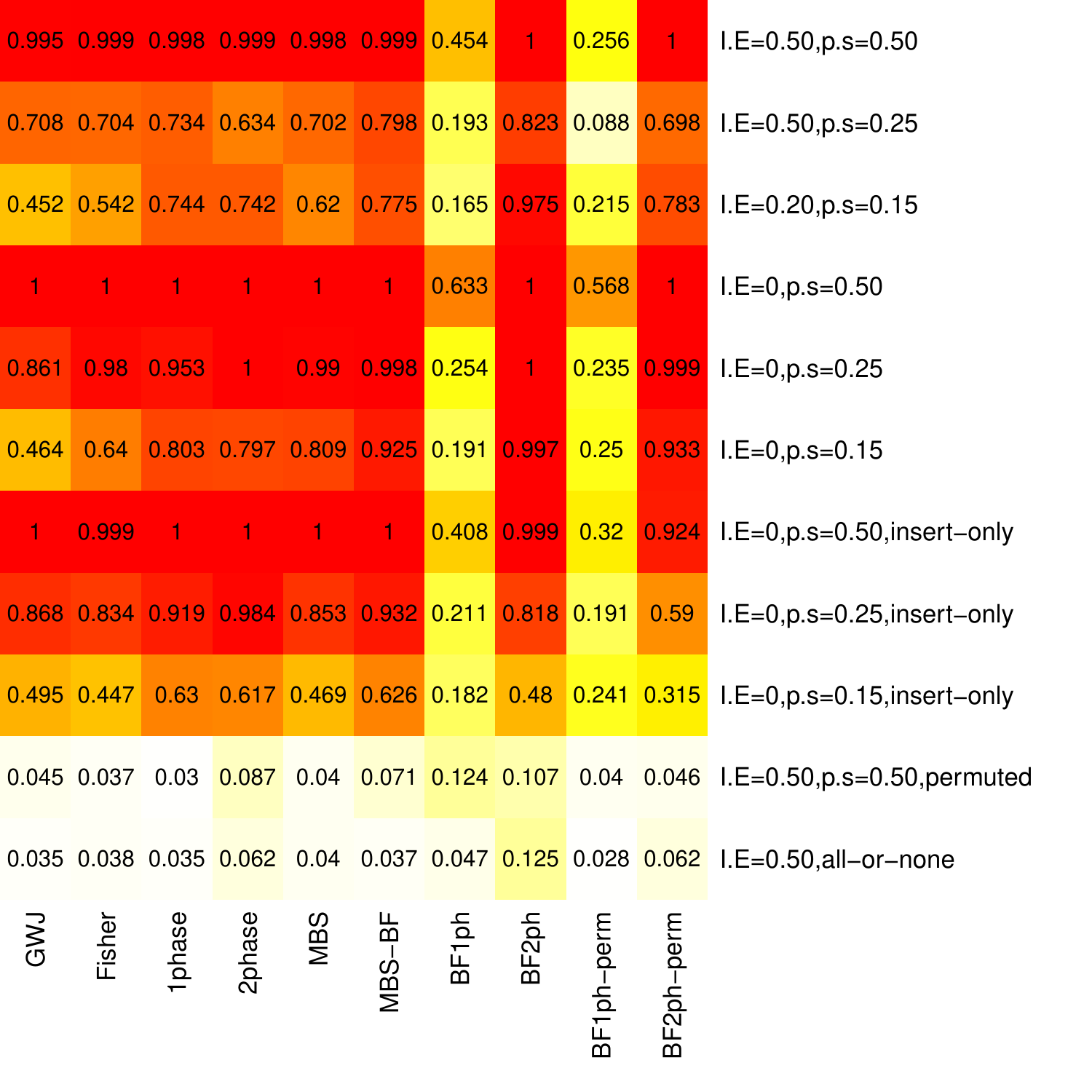}
\end{center}
\caption{Simulation study rejection rates heatmap, by scenario (row)
  and method (column).  These methods and scenarios are described in the main text. Numbers in cells indicate the fraction of the
  $1000$ scenario simulated data sets for which the method's
  decision rule favored the hypothesis of a sieve effect over no sieve
effect.  Cells are shaded or colored, with higher values in
darker-colored cells. As all frequentist tests are conducted at size $\alpha
= 0.05$, the values in the bottom two rows (the null-distributed data
scenarios) are at or near $0.05$ except for the Bayesian decisions
(``MBS'', ``BF1ph'', and ``BF2ph''.}
\end{figure*}

As depicted in Figure~\ref{f:heatmap}, we found that using analytic
likelihood ratio tests to compare the one-or-none model to the
all-or-none null model controls size effectively (even in $I_E>0$
cases in which Wilks's theorem does not apply), is powerful for
detecting weak sieve effects, and is robust to variations in the
assumptions governing the generating model (non-replacement-only,
replacement-only, and replacement-only insert-only).  The
replacement-only scenarios are indicated by ``I.E=0''.

We also found that for maximum-likelihood methods the two-phase approach performed comparably to
the one-phase approach, and for the Bayesian methods it performed much
better (a point we will address further below).

Indeed the only method with competitive power to the one-phase analytic-null
approach in these simulations is the ``MBS-BF'' method, which in these
examples also controls size reasonably well, but in general is not guaranteed
to do so.  Note that in these simulations, MBS (and MBS-BF)
perform well for scenarios with $I_E>0$, despite employing only
replacement-only models.

\begin{figure*}
  \label{f:ROC}
\begin{center}
\includegraphics{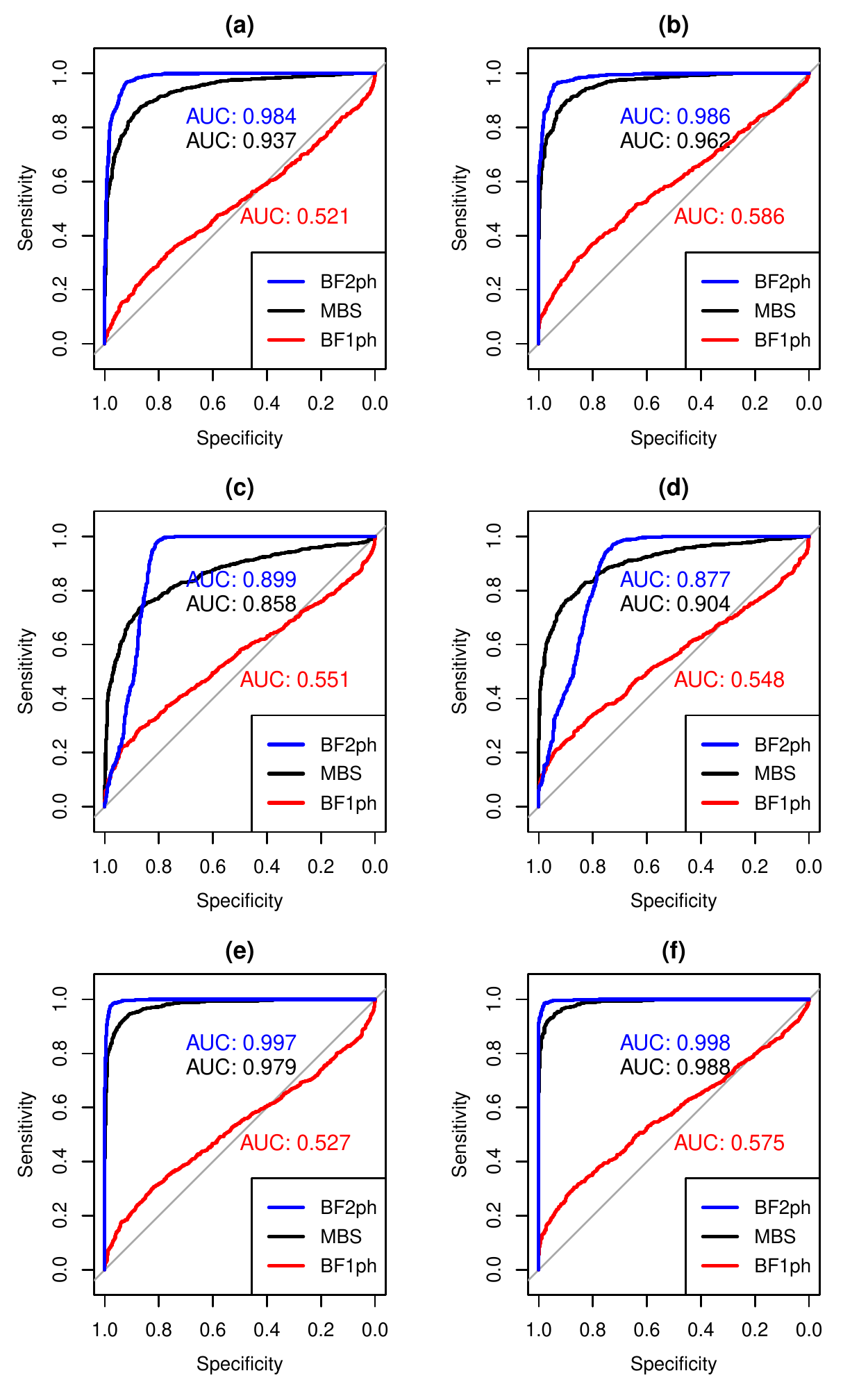}
\caption{ROC Curves comparing Bayesian testing methods, for scenarios with $p_s = 0.15$.  Figures in the left column (a, c, and e) use the permuted null scenario data as the true-negatives, whereas figures in the right column (b, d, and f) use the all-or-none null data.  Figures in the first row (a, b) use the $I_E = 0.2$ (non-replacement-only) scenario data as the true-positives, whereas figures in the second (c, d) and third (e, f) rows use the insert-only and replacement-only $I_E = 0$ scenario data, respectively.}
\end{center}
\end{figure*}

In these simulation experiments we also found that a simple test based
on Bayes factors (rejecting the null hypothesis whenever the BF favors the
one-or-none model over the all-or-none null model) performed
consistently well, both with the one-phase and two-phase models
(``BF1ph'' and ``BF2ph'') and with the MBS replacement-only mixture
model (``MBS-BF'').  However, this procedure is not guaranteed to
control size at a nominal rate.  A more apt evaluation of the Bayesian
methods, which assesses the area under the ROC curves, shows that over the range of thresholds on the Bayes
factors, the some-or-none two-phase approach outperforms the MBS-BF
approach for non-replacement-only scenarios, as shown in Figure~\ref{f:ROC}. Variations of these methods (``BF1ph-perm'', ``BF2ph-perm'', and ``MBS'', respectively), that
compare the data-estimated Bayes factors to null distributions
estimated by permutation of the class labels, are guaranteed to control
size (for that fixed-margin null), with some sacrifice of power.

For this example data, the Fisher's exact test performed well for
detecting moderate-sized sieve effects.  The Gilbert, Wu, Jobes
non-parametric t-test, evaluated using the HIV-b substitution matrix,
performed relatively poorly in comparison to the other methods,
particularly for the scenarios in which the non-insert distribution
matters (all but the ``insert-only'' scenarios).  GWJ performs
particularly well for the insert-only scenarios because the HIV-b
weight matrix strongly differentiates the target category from the
untargeted categories, but doesn't differentiate much among the
untargeted categories (so the GWJ test with this matrix approximates a
dichotomized Fisher test, little influenced by the distribution of the
untargeted categories).

The new approaches outshine GWJ and Fisher for low $p_s$ values (see
$p_s = 0.15$ results for $I_E = 0$ and $I_E = 0.2$) when the
non-insert distribution matters.  In practice, the new methods can be
seen as a complement to the routinely-used GWJ method, and can aptly
be described as being more sensitive than GWJ to differences among
non-targeted categories.


\subsection{STEP}

The sieve analysis of the breakthrough infections in the STEP (HVTN
502) HIV-1 vaccine trial identified a strong sieve effect at site Gag
84 \citep{Rolland}.  
Because the STEP vaccine was designed to elicit T cell responses that
target HIV-1 after infection, it is reasonable in this case to
consider a replacement-only model.  Since the estimated vaccine
efficacy was negative, this analysis should be viewed with caution as the
SNL ``no-harm'' condition may not hold.  The upcoming sieve analysis
of the HVTN 505 HIV-1 vaccine efficacy trial,
which was recently halted to new enrollments due to a finding of efficacy
futility, will be a better example of a zero-efficacy setting.  We
present this analysis of HVTN 502 infections as a demonstration of how the SNL
methods could be applied to HVTN 505 or similar settings.

Table~\ref{t:STEP.gag84.and.RV144.env169.failureTables} depicts the failure types
observed at site Gag 84 among STEP participants who experienced HIV-1
infections during the trial. The sieve effect is so strong that power differences among
testing procedures are not relevant.  We can easily detect this
effect, for instance, with a Fisher's exact test (p = 0.001).


\begin{table}[ht]
\centering
\caption{STEP failure types at site Gag 84 and RV144 non-failure
  counts (labeled ``0'') and failure types at site Env 169}
\begin{tabular}{rrrrrrrrrrr}
\hline
    & \multicolumn{2}{c}{Gag 84} & & \multicolumn{7}{c}{Env 169} \\
\hline
 & T & V &  & 0 & K & Q & R & E & T & V \\ 
  \hline
V &   8 &  31 &  & 7909 &  30 &   9 &   2 &   1 &   1 &   1 \\ 
  P &  17 &   9 &  & 7914 &  57 &   7 &   2 &   0 &   0 &   0 \\ 
   \hline

\end{tabular}
\label{t:STEP.gag84.and.RV144.env169.failureTables}
\end{table}



Investigating these data with a two-phase analysis comparing the
one-or-none model to the all-or-none model, we compute the log-likelihood ratio as
-16.5, which has an (analytic) p-value of
9.3e-09, strongly supporting the one-or-none model over the
all-or-none model at this site.  To take a more conservative approach, we permute the treatment labels
as described above, to estimate a null distribution that conditions on
the marginal totals.  Since none of these $1000$ permuted-data -2*LLRs was as large as the
observed value of 32.99, we estimate the p-value as $p < 0.001$, again supporting a
sieve effect.

\begin{figure}
\begin{center}
\label{f:Gag84posterior}
\includegraphics{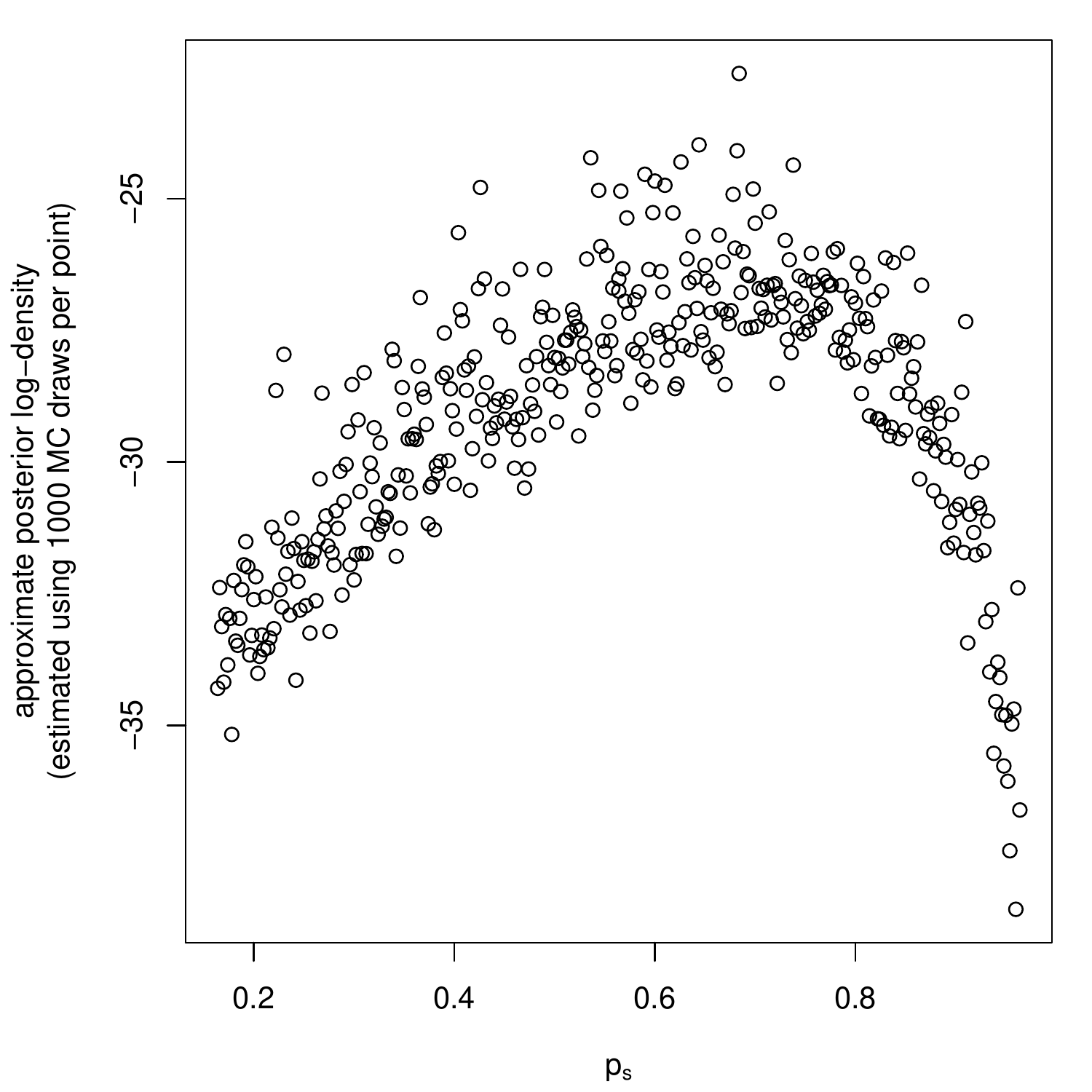}
\end{center}
\caption{Two-phase hierarchical one-or-none evaluation of STEP site
  Gag 84: The approximate posterior distribution of $p_s$, evaluated
  using 1000 Monte Carlo simulations per point.  The Y axis is
  log-scaled and unnormalized.  The posterior mass is concentrated
  well above zero, which since $I_E = 0$ indicates evidence for a sieve effect.}
\end{figure}

Now we turn to a Bayesian approach.  Figure~\ref{f:Gag84posterior} shows the approximate posterior
distribution of $p_s$ at this site when employing a hierarchical two-phase Bayesian approach (in which we use the
placebo-recipient data to update priors on both $\bm{q}$
and $\bm{p_c}$). The posterior-maximizing value of $p_s$ is
0.68, which we could interpret as indicating
that about two-thirds of the intervention recipients who became HIV-1
infected experienced immune escape at this locus that they would not
have experienced in the absence of the intervention.  Since this
analysis depends on the no-harm assumption, this interpretation should
be made cautiously.



\subsection{RV144}
The sieve analysis of the breakthrough infections in the RV144 HIV-1
vaccine trial identified a moderate sieve effect at site Env
169 \citep{v2sieve}.  The trial had an estimated intervention
efficacy of 31\% overall, so unlike the STEP analysis, here we need
not assume a replacement-only model.  Table~\ref{t:STEP.gag84.and.RV144.env169.failureTables} depicts the failure types
observed at site Env 169 among the 109 RV144 participants who experienced 
infections during the trial by subtype AE (CRF-01) HIV-1 viruses.  The effect is weaker at this site, so power differences between
testing procedures are relevant.  We would not detect this
effect, for instance, with a Fisher's exact test (p = 0.089).

We now investigate this site using the dichotomous-take SNL
framework, with a two-phase analysis
comparing the one-or-none model to the all-or-none model.  We compute the two-phase log-likelihood ratio as
-31.95, which has an (asymptotic) p-value of
1.3e-15, indicating strong evidence supporting a sieve effect. Since $I_E>0$, we do not expect Wilks's theorem to apply under the
null.  To take a more conservative approach, we permute the treatment
labels to estimate a null distribution.  Three of the $1000$
permuted-data $-2 \otimes$LLRs were larger than the observed value of
63.9, so this
result is much less significant than when employing the chi-squared null.

\begin{figure}
\begin{center}
\label{f:Env169posterior}
\includegraphics{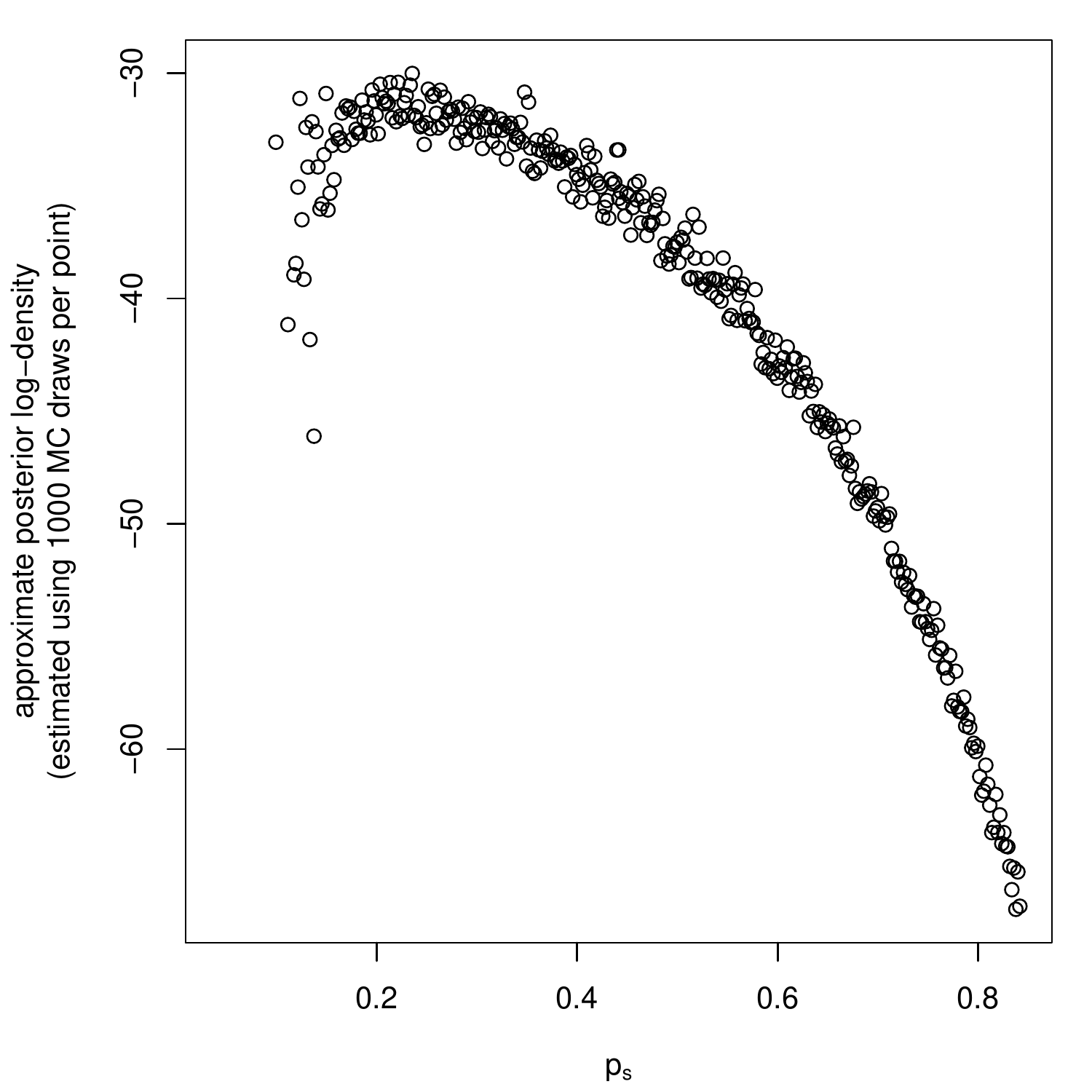}
\end{center}
\caption{Two-phase hierarchical one-or-none evaluation of RV144 site Env 169: The approximate posterior distribution of $p_s$, evaluated
  using 1000 Monte Carlo simulations per point.  The Y axis is
  log-scaled and unnormalized.  The posterior mass is concentrated
  near $p_s = 0.24$, indicating that about that fraction of the
  partial efficacy of RV144 can be attributed to the sieve effect.}
\end{figure}

Now we turn to a Bayesian approach.  We
estimate the hierarchical (for both $\bm{p_c}$ and $\bm{q}$) two-phase
Bayes factor at this site to be
3.1e+10,
which is much greater than $1$ and thus supports the one-or-none model
over the all-or-none model.  In Figure~\ref{f:Env169posterior} we
investigate the posterior distribution of $p_s$ at RV144 site Env 169.

The posterior-maximizing value of $p_s$ is 0.24, which we interpret as indicating that about one-quarter of
the intervention recipients who became HIV-1 infected would not have
become infected if the intervention had targeted all amino acids at site Env 169 rather than just the insert AA.  That is, the HIV-1 infecting these subjects was able to escape the vaccine-induced immune pressure because the intervention targeted just the K, not the other AAs, at site Env 169.  This analysis suggests that the incompleteness of the efficacy can only partly be attributed to the sieve effect.  In this non-replacement-only case, the sieve effect strength is 
not simply equivalent to the rate of ``take''; as discussed in section
\ref{s:p_s}, the interpretation is that incomplete ``take'' accounts
for $1-p_s$ (about three-quarters) of the unavoided vaccine-recipient
failures, indicating that even if the vaccine had induced (in takers)
immune responses that successfully targeted all amino acids at Env
169, the vaccine would still have been only partially efficacious
(efficacy would have been about 50\% rather than 31\%).

\section{Discussion}
\label{s:discuss}
%

We have introduced a framework for sieve analysis of
non-leaky interventions that do not cause harm (the intervention induces
only ``failure avoidance'', not new failures except as replacements of
avoided failures).  The framework provides a useful platform for
reasoning about sieve effects, and we argue that apparent sieve
effects are not generally attributable to actual differential efficacy
across failure types except under either the ``leaky''
conditions of \cite{GSA}, or under the
no-harm, non-leaky conditions of this framework.

We have introduced models for evaluating sieve
effects in this sievey-not-leaky setting.  We have implemented both frequentist
and Bayesian procedures, and we have demonstrated the power of
both frequentist and Bayesian-frequentist hybrid approaches under
various conditions.  The primary parameter of the model, the ``sieve effect strength''
$p_s$, has the useful interpretation that it reflects the amount of an
intervention's partial efficacy that can be attributed to the sieve
effect (as opposed to incomplete ``take'' of the intervention by some
intervention recipients).

For replacement-only ($I_E=0$) scenarios, since the all-or-none model is
nested in any some-or-none model whenever $p_s = 0$, the hypothesis
test based on likelihood ratios has a simple asymptotic
analytic form.  We found in our simulation experiments that the same
analytic approach works well for scenarios with $I_E > 0$ (to which the
asymptotics do not apply).

The SNL framework that we have introduced is
generalizable beyond the dichotomous ``take'' assumptions that we have
employed here.  At the extreme, we could allow non-zero probabilities for
all of the $2^J$ possible response types allowed by the some-or-none
framework (these index the set of types that are targeted, including
both ``all'' and ``none'').  One particular example would constrain
these to ensure independent ``take'' probabilities for each targeted type.  It
remains for future work to develop models and tests for this scenario.

With now two sieve analysis frameworks for two extremes of the
spectrum (leaky interventions with 100\%
``take'' and this new approach for non-leaky interventions with
partial ``take''), there remains a need for development of methods that
support unbiased evaluation of sieve effects under conditions of both
leaky and incomplete-take interventions.  In the mean time, or if that
is not possible, there remains a need for further development of tests
to evaluate the conditions under which these methods yield reliable
results.

We conclude with a brief discussion of the ``no-harm'' assumption.  Mathematically, the no-harm
condition requires that the per-exposure probabilities of infection are
always $1$ under the control (or counterfactual or no-take)
intervention scenario.  This means that ``exposure'' 
is equated with ``an exposure that would result in infection in the
absence of the treatment intervention.'' Here, the
definition of ``exposure'' reflects a modeling choice that determines
the tradeoff between the per-exposure probabilities and the exposure distributions.

For partially effective
interventions, the choice may be arbitrary, but it is clearly not a reasonable assumption for an
intervention with a significantly negative estimated
efficacy.  In such scenarios, it may still be possible to use the
sievey-not-leaky failure avoidance framework by considering the control intervention an
alternative treatment, and supposing a shared set of would-be-first
failures (in a no-harm setting) that are avoided at different rates by
the two interventions.  We leave this to future work.


\backmatter


\section*{Acknowledgements}

The author thanks Peter Gilbert for helpful guidance and critical reading of the paper. This research was supported by NIH NIAID grant 2 R37 AI054165-11. The author thanks the participants, investigators, and sponsors of the RV144 trial, including the U.S. Military HIV Research Program (MHRP); U.S. Army Medical
Research and Materiel Command; NIAID; U.S. and Thai
Components, Armed Forces Research Institute of Medical Science;
Ministry of Public Health, Thailand; Mahidol University; SanofiPasteur; and
Global Solutions for Infectious Diseases. The author thanks the participants, investigators, and sponsor of the STEP trial, the NIH NIAID.  The opinions expressed in this Article are those of the author and do not represent the official views of the NIAID.\vspace*{-8pt}


\section*{Supplementary Materials}

Web Appendices A, B, and C, referenced in Section~\ref{s:p_s}, as well
as Web Appendix D, referenced in Section~\ref{s:simulation}, are available with
this paper at the Biometrics website on Wiley Online
Library.\vspace*{-8pt}




%
%
%

\appendix


\section{}

\subsection{Web Appendix A: The sieve effect strength $p_s$}

Here we show that the ``sieve effect strength'' $p_s$ expresses the magnitude of a sieve
effect as the proportion of those subjects who would not be protected by
the some-or-none intervention who would instead be protected if the
intervention were an all-or-none intervention.  

Note that an all-or-none intervention has the property that for all failure types $j \in 1, \dotsc, J$,
$p_{vj} = p_{cj}$.  If the intervention is effective (ie if $p_t > 0$)
then all responders who would have failed are instead protected, so
$r_{v0} = r_{c0} + ( 1 - r_{c0} )p_t$, and the intervention reduces
the rates of all failures by the same amount, $p_t$.

In contrast, only a subset of the responders would be protected by a some-or-none
intervention: only the subjects who would have failed with a
targeted-type failure in the absence of any intervention.  Also, in
order to be protected a subject must have no (untargeted) replacement failures.
Thus for a one-or-none intervention, $r_{v0} = r_{c0} + r_{c1} p_t ( 1
- p_2 )$.  For a some-or-none intervention, with $\omega_{1j} = 1$
for categories $j \in {1, \dotsc, g}$ and zero otherwise, $r_{v0} = r_{c0} +
r_{cG} p_t ( 1 - p_2 )$.

For some-or-none models with $g > 1$, both $p_{vG} = p_{cG}( 1 - p_s
)$ (overall) and $p_{vj} = p_{cj}( 1 - p_s )$ (for each $j \in 1,
\dotsc, g$).  This derives from $p_{vG} = p_{cG}( 1 - p_s )$, since
$p_{vj} = p_{vG}\frac{p_{cj}}{p_{cG}}$ so $p_{vj} =
\frac{p_{cj}p_{cG}( 1 - p_s )}{p_{cG}} = p_{cj}( 1 - p_s )$.  This
also means that simultaneously, $p_s = 1 - \frac{p_{vG}}{p_{cG}}$
(overall) and $p_s = 1 - \frac{p_{vj}}{p_{cj}}$ (for each $j \in 1,
\dotsc, g$), and therefore that the relative rate of type-specific
intervention-associated reduction in failures is constant across the
targeted categories.  This is consistent with the SNL sieve condition that
ensures that a subject's probability of having response $r_i = 1$ is
independent of the rates of her would-be-first failure processes.

The failure type target(s) of a one-or-none or some-or-none
intervention will be affected differently than the off-target
categories.
For some-or-none interventions targeting types $1$ through
$g$, $r_{vj} = r_{cj}( 1 - p_t
)$ for all $j \in 1, \dotsc, g$ and $p_s \equiv \frac{p_t ( 1 - p_{cG}( 1 -
  p_2 ) )}{1 - p_t p_{cG} (1 - p_2 )}$.

To see that the sieve effect strength reflects the amount of an intervention's partial efficacy
that can be attributed to the sieve effect, consider
first that its denominator is $1 - I_E$, where $I_E$ is the
intervention's efficacy as defined by the fraction of the subjects
that would have failed in the absence of the intervention who did not
fail in the presence of it:
$$
I_E = 1 - \frac{n_v - n_{v0}}{n_c - n_{c0}} = \frac{n_{cG}p_t(1-p_2)}{n_c - n_{c0}} = p_{cG} p_t ( 1 - p_2 ).
$$

Another interpretation of the denominator $1 - p_t
p_{cG} (1 - p_2)$ of the sieve effect strength
$p_s$ is that it is the probability, for a person who would have
failed in the absence of any intervention, of that person still
failing despite being given the some-or-none intervention.  The additive inverse of the sieve effect strength
parameter is $(1 - p_s) = \frac{1-p_t}{1 - p_t p_{cG} (1 - p_2)}$,
which is the probability that a subject is not a responder ($1-p_t$) divided by the
probability that a failure would not be avoided in the setting of a
some-or-none intervention.  The all-or-none intervention
protects all responders, so $(1 - p_s)$ is the ratio of failures
not avoided by the two intervention types (all-or-none
vs. some-or-none), and the sieve effect strength is the fraction of
those not protected by the some-or-none intervention whose
non-protection is attributable to the some-type versus all-type
coverage (that is, attributable to the sieve effect).  The rest of the
non-protection is attributable to incomplete take (ie. the fact that
$p_t < 1$) and would be experienced with both types of intervention.



\subsection{Web Appendix B: General constraints implied by the some-or-none model}
In general, $I_E \le p_t p_{cG}$ for all
some-or-none models, with or without replacement failures, and thus
(since $p_t \le 1$), $I_E \le p_{cG}$.  Circumstances in which $I_E > p_{cG}$ are unlikely to be well-modeled
by dichotomous-take non-leaky some-or-none failure avoidance models targeting types
$1, \dotsc, g$.  This could in principle lead to a pre-screening procedure to
exclude from consideration any some-or-none model in significant
violation of the condition that $I_E \le p_{cG}$.  We leave details to
future work.

Each of the parameters of the some-or-none model is directly
estimable from the data, given the other parameters.  By plugging-in estimates $\hat{I_E}$, $\hat{p_{vj}}$,
  and $\hat{p_{pj}}$ of $I_E$, $p_{vj}$ and
$p_{cj}$ for $j \in 1, \dotsc, J$, the sieve effect strength can be estimated by $\hat{p_s} = 1 -
 \frac{\hat{p_{vG}}}{\hat{p_{pG}}}$, the marginal response rate by $\hat{p_t}
 = 1 - \frac{1 - \hat{p_s}}{1 - \hat{I_E}}$, and the cause replacement rate by $\hat{p_2} = 1 -
 \frac{\hat{I_E}}{\hat{p_{cG}} \hat{p_t}}$.  Even the conditional
 category replacement distribution can be estimated, since for
 non-targeted types, $\hat{q_j} = \frac{(1-\hat{I_E})\hat{p_{vj}} -
   \hat{p_{cj}}}{\hat{p_{cG}} \hat{p_t} \hat{p_2}}$.  These plug-in estimates
 will not always satisfy all of the simultaneous constraints,
 however, such as the constraint that $\sum{\hat{q_j}} = 1$ or the
 constraints that individual parameters are valid
 probabilities (eg $\hat{p_2}$ must satisfy $0 \le \hat{p_2} \le 1$).

 All required constraints will be satisfied under the bounds imposed by our
 parameterization if (and only if) the following relation holds: $I_E \le p_{cG} p_t$.
 This constraint can be equivalently expressed as a one-sided boundary
 on any one value in the set $\{ I_E, p_{cG}, p_s \}$, given the other
 two values:
\begin{equation}
\begin{split}
   I_E &\le \frac{ p_{cG} p_s }{ 1 - p_{cG} ( 1 - p_s ) } \hbox{; and} \\
   p_{cG} &\ge \frac{I_E}{p_s + I_E ( 1 - p_s )} \hbox{; and} \\
   p_s &\ge \frac{I_E ( 1 -  p_{cG} )}{p_{cG} ( 1 - I_E )} \hbox{,}
\end{split}
\end{equation}
but often in practice we enforce the constraint via the alternate
equivalent expression, $I_E \le p_{cG} ( 1 - ( ( 1 - p_s ) ( 1 - I_E
) ) )$.

Note that this constraint implies that $I_E \le p_{cG}$, since $p_t \le 1$.

In some settings additional constraints arise,
such as when $p_2$ is assumed equal to zero (under the assumption of
no replacement failures), which implies the additional constraint that
$p_{vG} = \frac{p_{cG} - I_E}{1 - I_E}$, or when $p_2$ is assumed equal to one (under the
replacement-only model), which implies the additional constraint that
$I_E = 0$.  We sometimes specify (rather than estimate) the
conditional replacement type distribution $q_j$ for $j > g$, which
also constrains the other parameter estimates.

There will not generally be a some-or-none model that fits the
naive plug-in estimates (such that all the constraints are satisfied). The
some-or-none model with maximum likelihood will employ parameter estimates
that deviate from the naive MLEs.  One
strategy, discussed in the main text (the two-phase non-hierarchical approach), is to perform the some-or-none model fit
under conditions of fixed estimates of $I_E$, $r_{c0}$ and the
vector of $p_{cj}$ values.  In these
circumstances the model's constraints restrict the estimates of the
values of $p_{vj}$ through parameters $p_s$ and $q_j$.


When $p_s = 0$, all replacement-only some-or-none models become equivalent to the
replacement-only all-or-none model (with $I_E=0$).

In the replacement-only some-or-none model, setting $p_s = 0$ forces
$p_t = I_E = 0$ (since
$p_s = 1 - \frac{1 - p_t}{1 - I_E}$).  The all-or-none model also has
$p_t = I_E$, since (in that model)
\begin{equation}
\begin{split}
  I_E &= 1 - (1-r_{v0})/(1-r_{c0}) \\
      &= 1 - (1-(r_{c0}+r_{cG}p_t))/(1-r_{c0}) \\
      &= \frac{(1-r_{c0}) - (1-(r_{c0}+r_{cG}p_t))}{1-r_{c0}} \\
      &= \frac{r_{cG}p_t}{1-r_{c0}} = p_t p_{cG} \hbox{,}
\end{split}
\end{equation}
and $p_{cG} = 1$.

\subsection{Web Appendix C: Some-or-none model likelihood}

We express the likelihood in terms of the primary parameters,
$\bmath{p_c} = \{p_{cj}: j \in 1, \dotsc, J\}$, $p_s$, $I_E$,
$r_{c0}$, and $\bmath{q} = \{ q_j: j \in (g+1), \dotsc, J
\}$. $r_{c0}$ is included because $I_E = 1 - \frac{1 - r_{v0}}{1 -
  r_{c0}}$, and the full likelihood involves 
$$
Pr( n_{p0}, n_{v0} | r_{c0}, r_{v0}, n_p, n_v ) = P_b( n_{p0} | n_p, r_{c0}) P_b( n_{v0} | n_v, r_{v0}) \hbox{,}
$$
where $P_b(x|n, p)$ represents the probability of $X=x$ when $X$ is a
random variable following a binomial law with parameters $n$ and $p$.
The rest of the likelihood does not depend on $r_{v0}$ and $r_{c0}$
except through $I_E$.  Therefore, when $I_E$ is known (or assumed
equal to $0$ as in the ``replacement-only'' models), this contribution
from the non-failing subject count data to the overall likelihood cancels in the context of a likelihood ratio statistic or a Bayes factor.

Here we present the full joint likelihood for a some-or-none model
under the conditions of the dichotomous-take non-leaky failure avoidance ``SNL'' framework, which (for a
some-or-none intervention targeting types $1, \dotsc, g$) is given by 
\begin{equation}\label{eq:full.likelihood.someornone.0}
\begin{split} 
Pr( \bmath{Y} &| M_a^{g-or-none}, \bmath{p_c}, p_s, I_E, r_{c0}, \bmath{q} ) =\\
&Pr( n_{p0}, n_{v0} | I_E, r_{c0}, n_p, n_v ) \\
&\times P_m( \bm{n_p} | \bmath{p_c} ) { P_m( \bm{n_v} = \{ n_{vj}: j \in 1, \dotsc, J \} | \bm{p_v} = \{ p_{vj} : j \in 1, \dotsc, J\} ) } \hbox{,}
\end{split}
\end{equation}
where
\begin{equation}
\begin{split} 
p_{vj} &= p_{cj}( 1 - p_s )I( j \in 1, \dotsc, g ) \\
&\; + \frac{p_{cj} + p_{cG} p_t p_2 q_j}{1 - I_E} I( j \in (g+1), \dotsc, J) \hbox{,}
\end{split}
\end{equation}
and where $P_m( \bm{x} | \bm{p} )$ represents the probability of counts $\bm{x}$ under a multinomial model with parameters $n=\sum{ x_j }$ and $\bm{p}$.
    
Note that since $p_2 = 1 - \frac{I_E}{p_{cG}p_t} = \frac{p_{cG}p_t -
  I_E}{p_{cG}p_t}$, $p_{cG} p_2 p_t = p_{cG} p_t - I_E$.  Also, since
$1 - p_s = \frac{1 - p_t}{1 - I_E}$, $p_t = 1 - ( 1 - p_s )( 1 - I_E
)$.  Thus $p_{cG} p_2 p_t = p_{cG} ( 1 - ( 1 - p_s )( 1 - I_E ) ) -
I_E$.  Then $p_{vj}$ in the likelihood can be expressed solely in terms of the primary parameters as
\begin{equation}
\begin{split} 
  p_{vj} &= p_{cj}( 1 - p_s )I( j \in 1, \dotsc, g ) \\
  &\; + \frac{p_{cj} + ( p_{cG} ( 1 - ( 1 - p_s )( 1 - I_E ) ) - I_E )
    q_j}{1 - I_E} I( j \in (g+1), \dotsc, J ) \hbox{.}
\end{split}
\end{equation}

\subsection{Web Appendix D: Equivalence of the three insert-only models}
There are three types of insert-only model: either $p_2 = 0$, in which
case we
call it the ``no replacement insert-only model'', or $\bm{q} =
\bm{q_c}$, which we further
differentiate by whether $I_E = 0$, in which case we call it the
``replacement-only insert-only model'', or $I_E > 0$, in which case we
call it the ``non-replacement-only insert-only model''.  Here we show
that all three of these are equivalent in their expressions of the
category probabilities $\bm{p_v}$.

Note that these models are not exactly equivalent.  For example, in
the ``no replacement insert-only'', $p_s = \frac{p_t ( 1 - p_{c1} )}{1 - p_t
  p_{c1}}$, whereas in the ``replacement-only insert-only''
model, $p_s = p_t$, and for the ``non-replacement-only insert-only''
model, $p_s \equiv \frac{p_t ( 1 - p_{cG}( 1 - p_2 ) )}{1 - p_t p_{cG} (1 - p_2 )}$.

However, they all involve no intervention effect on the conditional
distribution among the non-targeted categories, since when $p_2 = 0$,
$q_{vj} = q_{cj}$ for all $j > g$.  Since also in all models, $p_{vj}
= p_{cj} ( 1 - p_s )$ for targeted types $j \in 1, \dotsc, g$, for
practical purposes when expressed using $p_s$, the models are equivalent.  Their only difference
is in interpretation of the sieve effect strength $p_s$ in relation to
$p_t$ and $I_E$.

In the setting of the replacement-only insert-only
model, $I_E = 0$ and $p_s = p_t$, but in the some-or-none model with
no replacement failures, $1 - p_s = \frac{1 - p_t}{1 - I_E} = 1 -
\frac{p_t ( 1 - p_{c1} )}{1 - p_t p_{c1}}$,
so $p_t = \frac{I_E}{p_{cG}}$, $p_s = \frac{I_E}{1 - I_E} \frac{1
  - p_{cG}}{p_{cG}}$, and $p_{v1} = p_{cG}( 1 - p_s ) = \frac{p_{cG} - I_E}{1 - I_E}$.


Thus in the ``no replacement insert-only'' model, the fraction of failures that
are of the targeted types is reduced by an amount proportional to the
intervention effect, whereas in the ``replacement-only insert-only''
model there is no intervention effect, and the fraction is instead
reduced by an amount proportional to the ``take'' response rate.  The
interpretation of $p_s$ for the ``non-replacement-only insert-only''
model is the same as for the general some-or-none model with arbitrary
replacement distribution $\bm{q}$.

\label{lastpage}

\end{document}